\documentclass[aps,preprint,superscriptaddress,nofootinbib]{revtex4-1}   
\pdfoutput=1
\usepackage[utf8]{inputenc}
\usepackage{amssymb}
\usepackage{amsmath}
\usepackage{amsfonts}
\usepackage{graphicx}
\usepackage{color}
\usepackage{xspace}
\usepackage{comment}
\usepackage{hyperref}
\usepackage[normalem]{ulem}

\usepackage[section]{placeins}
\usepackage{afterpage}

\usepackage{float}
\usepackage{slashed}
\usepackage{ulem}
\usepackage{appendix}

\usepackage{cancel}

\usepackage{multirow,rotating}
\usepackage[dvipsnames]{xcolor}
\usepackage{orcidlink}   

\begin{document}
	
      \title{Searching for apparent baryon number violation in $\Lambda_c^+$ decays at the Super Tau-Charm Facility}

	   \author{Zeren Simon Wang\,\orcidlink{0000-0002-1483-6314}}
	   \email{wzs@hfut.edu.cn}
      \affiliation{School of Physics, Hefei University of Technology, Hefei 230601, People’s Republic of China}

      \author{Xin-Ru Tang\,\orcidlink{0009-0005-9671-321X}}
      \email{tangxinru\_usc@163.com}
      \affiliation{University of South China, Hengyang 421001, People’s Republic of China}

      \author{Yu Zhang\,\orcidlink{0000-0001-9956-4890}}
      \email{yuzhang@usc.edu.cn}
      \affiliation{University of South China, Hengyang 421001, People’s Republic of China}

      \author{Yu Zhang\,\orcidlink{0000-0001-9415-8252}}
      \email{dayu@hfut.edu.cn}
      \affiliation{School of Physics, Hefei University of Technology, Hefei 230601, People’s Republic of China}

      \author{Xiaorong Zhou\,\orcidlink{0000-0002-7671-7644}}
      \email{zxrong@ustc.edu.cn}
      \affiliation{University of Science and Technology of China, Hefei 230026, People’s Republic of China}

	\begin{abstract}       
           Observation of baryon number violation (BNV) in laboratory experiments would constitute unambiguous evidence for physics beyond the Standard Model. We propose dedicated searches for \textit{apparent} BNV in charm-baryon decays, $\Lambda_c^+\to M^+ +$~missing energy ($M=\pi, K$) where the missing energy stems from a resonance. These channels have not been explored experimentally so far, despite the relatively clean environment potentially provided by near $\Lambda_c^+\overline{\Lambda}_c^-$ threshold production at $e^+e^-$ colliders. Performing state-of-the-art Monte Carlo simulations for the proposed Super Tau-Charm Facility (STCF), we evaluate the signal efficiencies and derive projected model-independent sensitivities under the assumption of negligible background. We further interpret these sensitivities within two theoretical frameworks: a sterile-neutrino-extended low-energy effective field theory ($\nu$LEFT) and R-parity-violating (RPV) supersymmetry. With an integrated luminosity of 1~ab$^{-1}$, STCF can probe new-physics scales of several TeV in the $\nu$LEFT description and constrain the RPV model parameter $\lambda''_{212}/m^2_{\tilde{q}}$ down to about $0.1~\mathrm{TeV}^{-2}$. Our results demonstrate that STCF provides a highly competitive opportunity for probing BNV interactions in rare charm-baryon decays.
 	\end{abstract}



	\maketitle
    \noindent

\section{introduction}\label{sec:intro}

Baryon number is conserved in the Standard Model (SM) at all orders in perturbation theory, and can be violated only through non-perturbative effects such as sphalerons~\cite{Klinkhamer:1984di} and instantons~\cite{Polyakov:1975rs,Belavin:1975fg,tHooft:1976rip,tHooft:1976snw}.
The non-perturbative effects are, however, strongly suppressed at low energies such as the collider scales~\cite{Ringwald:1989ee,Arnold:1987mh,Quiros:1999jp,Morrissey:2012db}.
Therefore, observation of baryon-number-violating (BNV) processes in laboratory experiments would be unambiguous evidence of physics beyond the SM (BSM)~\cite{Broussard:2025opd}.

On the other hand, we live in a Universe with matter-antimatter asymmetry~\cite{Planck:2018vyg}.
To realize baryogenesis~\cite{Dolgov:1991fr,Riotto:1999yt,Morrissey:2012db}, Sakharov conditions~\cite{Sakharov:1967dj} are required, including baryon number violation.
However, experimentally we have not observed any BNV processes, while increasingly more stringent bounds have been established; for instance, the lower bound on the proton lifetime has been raised to $\mathcal{O}(10^{34})$ years~\cite{ParticleDataGroup:2024cfk}.
Thus, probing BNV experimentally is closely connected to understanding the origin of matter.

In this work, we study effective field theories (EFTs) extended with sterile neutrinos $\nu_s$ where BNV operators involving a massive sterile neutrino, which is an SM-singlet, can be written down.
In the neutrino-extended SM effective field theory ($\nu$SMEFT)~\cite{delAguila:2008ir,Aparici:2009fh,Liao:2016qyd,Li:2021tsq}, all new degrees of freedom beyond the SM, except the sterile neutrinos, are postulated to be well above the electroweak scale, and such new-physics (NP) effects are encoded in so-called Wilson coefficients for each corresponding effective operator.\footnote{See also Ref.~\cite{Heeck:2026dmh} for a recent comprehensive study of BNV operators in the $(\nu)$SMEFT.}
When we work at lower energy scales, we move to the framework of neutrino-extended low-energy EFT ($\nu$LEFT)~\cite{Bischer:2019ttk,Chala:2020vqp,Li:2020lba,Li:2020wxi}, valid down to roughly the hadron scales.
Working further down the energy scale requires matching the $\nu$LEFT to Baryon Chiral Perturbation Theory (BChPT)~\cite{Claudson:1981gh,Jenkins:1991ne,Liao:2025vlj}.

We will focus on BNV operators with a sterile neutrino that may lead to \textit{apparent} BNV processes at colliders; concretely, if a very long-lived sterile neutrino is produced, it can escape the main detector appearing as missing energy and thus the visible final state violates the baryon number.\footnote{Long-lived sterile neutrinos can also lead to displaced-vertex (DV) signatures at colliders and beam-dump experiments; see, e.g., Refs.~\cite{Beltran:2025ilg,Gunther:2023vmz,Bolton:2025tqw,Bertholet:2025lar,Braat:2026wgm} for some recent phenomenological studies on DV searches for long-lived sterile neutrinos in the EFT frameworks.}
Such apparent BNV processes have been studied and searched for in meson decays, in the theoretical frameworks of, for example, $B$-mesogenesis~\cite{Elor:2018twp,Nelson:2019fln,Alonso-Alvarez:2019fym,Alonso-Alvarez:2021qfd,BaBar:2023rer,BaBar:2023dtq} and R-parity-violating supersymmetry (RPV-SUSY)~\cite{Dib:2022ppx,BaBar:2023dtq,Belle-II:2026tyb}.\footnote{See Refs.~\cite{Dreiner:1997uz,Allanach:2003eb,Barbier:2004ez,Mohapatra:2015fua} for some reviews on the RPV-SUSY.}
Both model scenarios predict $B^+$ decays into a proton plus a dark baryon and a light bino neutralino $\tilde{\chi}^0_1$, respectively; the dark baryon and the light bino neutralino are so long-lived that they escape the main detector.
Such models are well motivated for explaining issues in the SM such as matter-antimatter asymmetry, hierarchy problem, or the non-vanishing light-neutrino masses.
Further model scenarios relying on non-renormalizable operators that predict similar meson-decay or baryon-decay signatures were studied in e.g., Refs.~\cite{Davoudiasl:2010am,Baldes:2014rda,Baldes:2014gca,Heeck:2020nbq,Barman:2021tgt,Goudelis:2022bls,Ciscar-Monsalvatje:2023zkk,Li:2024liy,Blazek:2025wmc,Heeck:2025uwh,Fan:2025xhi,Helo:2025kgx,Bhoonah:2025qzd,Ma:2025mjy}.

We investigate the search prospect of apparent BNV in $\Lambda_c^+$ decays at the Super Tau--Charm Facility (STCF)~\cite{Achasov:2023gey,Ai:2025xop}, a facility proposed for operation in Hefei, China.
STCF is a symmetric electron--positron collider proposed to be operated with center-of-mass (COM) energies ranging between 2~GeV and 7~GeV, with a peak luminosity of $0.5\times 10^{35}$ cm$^{-2}$\,s$^{-1}$.
An exceptionally clean $e^+ e^-$ environment with high intensities, STCF is expected to produce copiously the SM $\tau$-leptons and charm hadrons.
In particular, the Born cross section of the $\Lambda_c^+\overline{\Lambda}_c^-$ pair production near the threshold at $\sqrt{s}=4.682$~GeV has been measured at BESIII to be $188.1\pm1.6\pm6.3$~pb~\cite{BESIII:2023rwv}, which would correspond to $\sim 1.88\times 10^{8}$ events for an integrated luminosity of 1~ab$^{-1}$ at STCF.
This allows for testing rare decays of the $\Lambda_c^+$ baryon, considering also the excellent tracking and particle-identification capabilities at the proposed facility.

Motivated by the considerations given above, we propose dedicated searches for $\Lambda_c^+\to M^+ +$~missing energy at STCF with $M=(\pi, K)$, and the missing energy may stem from a very long-lived fermion such as a sterile neutrino or a light bino neutralino that has escaped the main detector.
Concretely, at the COM energy $\sqrt{s}=4.682$~GeV, we tag the $\overline{\Lambda}_c^-$ baryon and require the $\Lambda_c^+$ baryon to undergo the signal decay (almost at rest), leading to the clean signature of a visible meson plus missing energy where the missing energy manifests itself as a peak at the mass of the escaping fermion.
With the state-of-the-art tool \textsc{OSCAR}~\cite{Huang:2023kog,Li:2024tuy,Ai:2024yqx}, a detector simulation framework for STCF based on \textsc{GEANT4}~\cite{GEANT4:2002zbu,Allison:2016lfl}, we perform Monte Carlo (MC) simulations to estimate the signal-event reconstruction efficiencies.
We then derive upper bounds on the branching ratio BR$(\Lambda_c^+\to M^+ +\text{~missing})$ as functions of the invariant mass of the missing energy.
We further interpret these sensitivity results in terms of the model parameters in the $\nu$LEFT with a sterile neutrino and in the RPV-SUSY with a light bino neutralino.

In particular, the light bino neutralino in the RPV-SUSY, as light as in the GeV-scale, is allowed by all constraints~\cite{Grifols:1988fw,Ellis:1988aa,Lau:1993vf,Dreiner:2003wh,Dreiner:2013tja,Profumo:2008yg,Dreiner:2011fp} as long as it is the lightest supersymmetric particle (LSP) and it decays, and can be reinterpreted within the neutrino-extended EFT frameworks~\cite{Gunther:2023vmz}.
This is because the light bino, similar to the sterile neutrino, is an SM-singlet, and the squark masses are stringently constrained to be above about 2~TeV~\cite{CMS:2019zmd,ATLAS:2020xgt}.
Such a light bino can be produced in $\Lambda_c^+ \to K^+ + \tilde{\chi}^0_1$ decays and appear as missing energy, if a single tiny RPV coupling $\lambda''_{212}$ is assumed to be non-zero while all other RPV couplings vanish.

We note that these proposed searches can be performed at BESIII~\cite{BESIII:2009fln} with similar analyses.
However, for $B$-factory experiments such as Belle~II~\cite{Belle-II:2010dht,Belle-II:2018jsg} and BaBar~\cite{BaBar:2001yhh,BaBar:2013byz}, no such near-threshold production mode of $\Lambda_c^+ \overline{\Lambda}_c^-$ exists and the environment is relatively busier; thus, the missing-mass techniques or double tagging cannot be reliably applied.

This paper is structured as follows.
In Sec.~\ref{sec:model} we introduce the theoretical frameworks we work with, i.e., the sterile neutrinos in the $\nu$LEFT and the light bino neutralino in the RPV-SUSY.
We also offer a brief discussion on the decay length of the long-lived sterile neutrino and light bino neutralino there.
We proceed, in Sec.~\ref{sec:exp}, to discuss the STCF experiment and detail our search analyses including the simulation procedures in \textsc{OSCAR} and the computation procedure of the sensitivity reach.
Then in Sec.~\ref{sec:results} we present our numerical results, and we conclude the work in Sec.~\ref{sec:conclusions}.

\section{Model frameworks}\label{sec:model}

In this section we introduce the theoretical frameworks we employ in this study.
We will start with the $\nu$LEFT framework where we list the effective dim-6 BNV operators we focus on.
We then present the matching relations from the $\nu$LEFT to the BChPT, and embed them into a phenomenological effective Lagrangian.
We will then provide the expressions for the decay widths of $\Lambda_c^+\to \pi^+/K^+ + \nu_s$.
In addition, we will detail the RPV-SUSY scenario with a light bino neutralino as the LSP produced in $\Lambda_c^+$ decays via an RPV coupling $\lambda''_{212}$, leading to the signal process $\Lambda_c^+\to K^++\tilde{\chi}^0_1$.
We show how to match this UV-completion to the EFT.
Both the sterile neutrino and the light bino neutralino are very long-lived, appearing as missing energy at STCF.

We will confine ourselves to the mass range roughly between $m_p$ and $m_{\Lambda_c^+}-m_{M^+}$, where $m_{p/\Lambda_c^+/M^+}$ labels the mass of proton, $\Lambda_c^+$ baryon, and the charged meson ($\pi^+$ or $K^+$).
For lower masses, too fast proton decay rates can be easily induced, and the upper mass reach is determined by the kinematic thresholds.
Note that if the missing particle has a mass quite close to $m_{\Lambda_c^+}-m_{M^+}$, the accompanying meson $M^+$ would not have sufficient energy for detection, largely impairing the event-reconstruction efficiencies.

\subsection{The $\nu$LEFT framework}\label{subsec:eft}

\begin{figure}[t]
      \centering
      \includegraphics[width=0.49\linewidth]{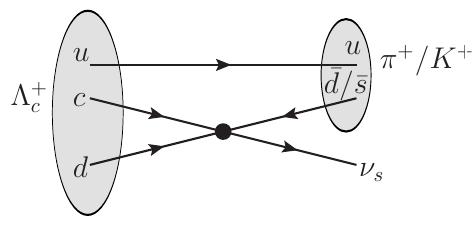}
      \caption{The parton-level Feynman diagram for the $\Lambda_c^+\to \pi^+/K^++\nu_s$ decays induced by effective operators. The black blob labels the effective vertex.}
      \label{fig:feynman_diagrams_eft_parton_level}
\end{figure}

Since we study $\Lambda_c^+$ decays in this work, we work with the $\nu$LEFT framework instead of the $\nu$SMEFT.
While multiple operators can lead to the signal processes, we choose to focus on the following ones involving the charm quark plus the down or strange quarks with right-chiral quark fields only, for an illustrative purpose:
\begin{eqnarray}
    \mathcal{O}^{S,RR}_{cdd}&=&\epsilon^{\alpha \beta \gamma}(\overline{c^c_{R\alpha}}d_{R\beta})(\overline{\nu_L}d_{R\gamma}),\label{eqn:ef_op_cdd}\\
    \mathcal{O}^{S,RR}_{cds}&=&\epsilon^{\alpha \beta \gamma}(\overline{c^c_{R\alpha}}d_{R\beta})(\overline{\nu_L}s_{R\gamma}),\label{eqn:ef_op_cds}\\
    \mathcal{O}^{S,RR}_{csd}&=&\epsilon^{\alpha \beta \gamma}(\overline{c^c_{R\alpha}}s_{R\beta})(\overline{\nu_L}d_{R\gamma}).\label{eqn:ef_op_csd}
\end{eqnarray}
Here, $\alpha, \beta, \gamma$ are color indices and $\epsilon^{\alpha\beta\gamma}$ is the anti-symmetric Levi-Civita tensor.
Following the standard $\nu$LEFT operator notation, the field $\nu_L$ denotes the left-handed chiral component of the sterile-neutrino field $\nu_s$, rather than an active SM neutrino.
We thus write down the dim-6 effective Lagrangian as
\begin{eqnarray}
    \mathcal{L}_{\text{eff.}}^{cdd}&\supset& \frac{c_{211}}{\Lambda^2}\mathcal{O}_{cdd}^{S,RR}+\text{h.c.},\label{eqn:cdd_eft_lag}\\
    \mathcal{L}_{\text{eff.}}^{cds,csd}&\supset& \frac{c_{212}}{\Lambda^2}\mathcal{O}_{cds}^{S,RR}+\frac{c_{221}}{\Lambda^2}\mathcal{O}_{csd}^{S,RR}+\text{h.c.},\label{eqn:cds_eft_lag}
\end{eqnarray}
where $c_{211}, c_{212}$, and $c_{221}$ are dimensionless Wilson coefficients and $\Lambda$ characterizes the corresponding NP scales.
Such EFT operators can induce our signal decays, $\Lambda_c^+\to M^+ + \nu_s$, as shown in Fig.~\ref{fig:feynman_diagrams_eft_parton_level}.

We note that in the present work, we focus on the right-handed operators $\mathcal{O}^{S,RR}$ as representative benchmark scenarios.
Similar apparent-BNV decays can also be induced by left-handed operators of the $qqd\nu_s$ type, generated by the $\nu$SMEFT operators $QQ\nu_s d$.
A phenomenological study of such operators has recently been presented in Ref.~\cite{Hiller:2026osz}.
We therefore restrict ourselves here to the complementary $udd \nu_s$-type operators.

We also comment that if the sterile neutrino mixes with active neutrinos, additional nucleon-decay constraints can arise.
In this case, the sterile-neutrino field appearing in the BNV operators can be connected to an active-neutrino component, and electroweak corrections may further induce standard BNV operators involving SM leptons.
Such effects are expected to be particularly relevant for operators containing left-handed quark doublets, since they couple directly to the electroweak gauge bosons.
For the right-handed $udd \nu_s$-type operators considered in this work, analogous effects require additional chirality flips or higher-loop corrections, and are therefore more model dependent and further suppressed.
A quantitative treatment of these mixing-induced nucleon-decay constraints is beyond the scope of the present work; here we assume that the sterile neutrino is sufficiently long-lived and has negligible active--sterile mixing, so that it appears as missing energy at STCF.

\begin{figure}[t]
      \centering
      \includegraphics[width=0.49\linewidth]{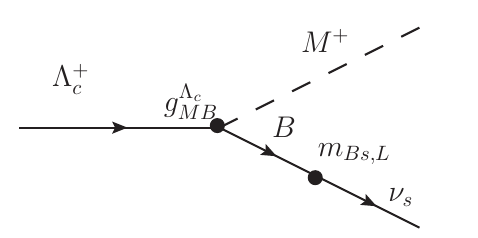}
      \includegraphics[width=0.49\linewidth]{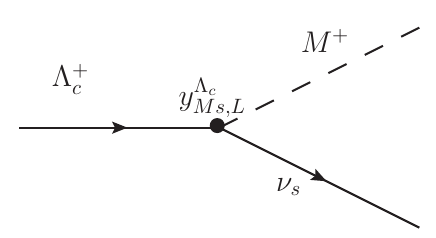}
      \caption{The hadron-level Feynman diagrams for the $\Lambda_c^+\to \pi^+/K^+ + \nu_s$ decays in the $\nu$LEFT framework. The left and right diagrams are for the ``pole'' and ``non-pole'' production mechanisms, respectively.}
      \label{fig:feynman_diagrams_eft_hadron_level}
\end{figure}

In order to compute the decay widths of $\Lambda_c^+\to \pi^+/K^+ + \nu_s$, we should first match the $\nu$LEFT to BChPT; for this purpose, we follow Ref.~\cite{Li:2025slp}, where BNV decays of the \textit{nucleons} are studied in the $\nu$LEFT. 
However, BChPT is rigorously applicable only to the light-flavor baryon octet, and its direct use for the charmed baryon $\Lambda_c^+$ lies outside its formal domain of validity.
To account for this mismatch, we promote the light-baryon-level (proton) matrix elements to the $\Lambda_c^+$ case and treat the corresponding hadronic form factor as an effective normalization parameter, to which we apply an order-unity variation.
Concretely, we vary the $\Lambda_c^+$ matrix-element form factor by multiplicative factors of $2$ and $1/2$.
We interpret this variation as reflecting the uncertainty associated with extrapolating the low-energy EFT matching into the heavy-flavor sector, thereby capturing the expected size of SU(3)-breaking and heavy-quark-symmetry-breaking effects in the absence of a controlled chiral expansion for charmed baryons.
This variation should therefore be understood as reflecting the intrinsic uncertainty of extrapolating BChPT beyond its regime of validity, rather than a controlled theoretical error.

We proceed to write down the $|\Delta(B-L)|=2$ effective Lagrangian for the interactions between a sterile neutrino, a meson, a baryon, and the $\Lambda_c^+$ baryon,
\begin{eqnarray}
    \mathcal{L}_\text{eff}=g^{\Lambda_c^+}_{MB}\overline{B}\gamma^\mu \gamma_5 \Lambda_c^+ \partial_\mu M+m_{Bs,L}\overline{\nu_s}P_R B+iy_{Ms,L}^{\Lambda_c^+}\overline{\nu_s}P_R \Lambda_c^+ M,
\end{eqnarray}
where the first two terms together lead to the $\Lambda_c^+ \to M^+  + \nu_s$ decays that proceed via an intermediate baryon $B$ that mixes with the sterile neutrino $\nu_s$, and the last term gives rise to direct production of $\nu_s$ in the same decay processes.
The two mechanisms are commonly dubbed as ``pole'' and ``non-pole'' contributions in the literature, respectively, and we illustrate them in Fig.~\ref{fig:feynman_diagrams_eft_hadron_level}.

We express the matrix element for the decay process $\Lambda_c^+\to M^+ + \nu_s$ as
\begin{eqnarray}
    i\mathcal{M}=\overline{u_{\nu_s}}P_R \Big( -y^{\Lambda_c^+}_{Ms,L} + \sum_B m_{Bs,L} \,\frac{\cancel{k}+m_B}{k^2-m_B^2}  \,  g^{\Lambda_c^+}_{MB} \, \cancel{p_M} \, \gamma_5\Big) \, u_{\Lambda_c^+},
\end{eqnarray}
where $p_M$ denotes the momentum of the outgoing meson.
We then compute the spin-averaged matrix element squared and obtain
\begin{eqnarray}
\overline{|\mathcal{M}|^2}/m_{\Lambda_c^+}^2&=&{1\over 2}|y_{Ms,L}^{\Lambda_c^+}|^2 (1+x_s^2-x_M^2)+\frac{1}{2}\sum_{B,B'} \,g_{MB}^{\Lambda_c^+} \, g_{MB'}^{\Lambda_c^+*} \, m_{Bs,L} \, m_{B's,L}^* \, g(x_B,x_{B'})\nonumber\\
&&-\sum_B {\rm Re}(y^{\Lambda_c^+}_{Ms,L} \, m_{Bs,L}^* \, g_{MB}^{\Lambda_c^+ *}) \, h(x_B)\;,
\label{eq:matrixelement}
\end{eqnarray}
where $x_s=m_{\nu_s}/m_{\Lambda_c^+}$, $x_M=m_{M^+}/m_{\Lambda_c^+}$, $x_{B^{(\prime)}}=m_{B^{(\prime)}}/m_{\Lambda_c^+}$, and
\begin{eqnarray}
g(x_1,x_2)&=&{(x_s^2+x_1x_2)(1-x_M^2-x_s^2(2+x_M^2)+x_s^4)-2x_s^2x_M^2(x_1+x_2)\over (x_1^2-x_s^2)(x_2^2-x_s^2)}\;,\\
h(x_B)&=&{x_B(1-x_M^2-x_s^2)-x_s^2(1+x_M^2-x_s^2)\over x_B^2-x_s^2}\;.
\end{eqnarray}
The decay width is given by
\begin{eqnarray}
\Gamma (\Lambda_c^+\to M^+ + \nu_s)=m_{\Lambda_c^+}{\lambda^{1/2}(1,x_M^2,x_s^2)\over 16\pi}{\overline{|\mathcal{M}|^2}\over m_{\Lambda_c^+}^2}\;,
\end{eqnarray}
where $\lambda(x,y,z)=x^2+y^2+z^2-2xy-2xz-2yz$.

\begin{table}[t]
\begin{tabular}{c|c|c|c}
Process          & $g^{\Lambda_c^+}_{MB}$          & $m_{Bs,L}$       & $y^{\Lambda_c^+}_{Ms,L}$      \\\hline
$\Lambda_c^+\to \pi^+ \nu_s$     & $g_{\pi \Sigma_c^0}^{\Lambda_c^+}=\frac{D+F}{f_\pi}$ & $m_{\Sigma_{cs,L}^0}=\frac{c_{211}}{\Lambda^2}\Big(\beta\Big)$ &    $y_{\pi s,L}^{\Lambda_c^+}=\frac{c_{211}}{\Lambda^2}\Big(-\frac{\beta}{f_{\pi}}\Big)$               \\\hline
\multirow{2}{*}{$\Lambda_c^+\to K^+ \nu_s$} &   $g_{K \Xi_c^{'0}}^{\Lambda_c^+}=\frac{D-F}{\sqrt{2}f_\pi}$       &     $m_{\Xi_{cs,L}^{'0}}=\frac{c_{221}}{\Lambda^2}\Big(\frac{\beta}{\sqrt{2}}\Big)$        & $y_{K s,L}^{\Lambda_c^+}=\frac{c_{212}}{\Lambda^2}\Big(-\frac{\beta}{f_{\pi}}\Big)$ \multirow{3}{*}{} \\
     &          $g_{K \Xi_c^{0}}^{\Lambda_c^+}=\frac{D+3F}{\sqrt{6}f_\pi}$            &    $m_{\Xi_{cs,L}^{0}}=\frac{c_{221}}{\Lambda^2}\Big(-\frac{\beta}{\sqrt{6}}\Big)$          &               \\
     &          &  \quad\quad\quad\quad $+\frac{c_{212}}{\Lambda^2}\Big(-\beta\sqrt{\frac{2}{3}}\Big)$                                                     &               
\end{tabular}
\caption{The derived effective couplings inducing the BNV decays of $\Lambda_c^+$ into a charged pion or kaon plus a sterile neutrino.}
\label{table:matching}
\end{table}

Refs.~\cite{Nath:2006ut,Beneito:2023xbk,Claudson:1981gh,Li:2025slp} derived the matching relations for light-flavor baryons and we, as discussed above, promote them to those for $\Lambda_c^+$, assuming that they take the identical forms.
We list these relations in Table~\ref{table:matching}, where the form factor $\beta_{\Lambda_c^+}=0.835\times 10^{-2}$~GeV$^3$ has been deduced in Ref.~\cite{Dib:2022ppx} from the results obtained with the approaches of QCD sum rules in Refs.~\cite{Azizi:2008ui,Wang:2009cr,Azizi:2015bxa,Azizi:2015tya}.
In the numerical analysis, we treat this result as a reference value for normalization and vary it by a factor of two.
In addition, the following values of the hadronic parameters $D$, $F$, and $f_\pi$ are used:
\begin{eqnarray}
 D=0.730~\text{\cite{Bali:2022qja}}, F=0.447~\text{\cite{Bali:2022qja}}, f_\pi=0.13041~\text{GeV}~\text{\cite{ParticleDataGroup:2024cfk}}.
\end{eqnarray}

\begin{figure}[t]
      \centering
      \includegraphics[width=0.7\linewidth]{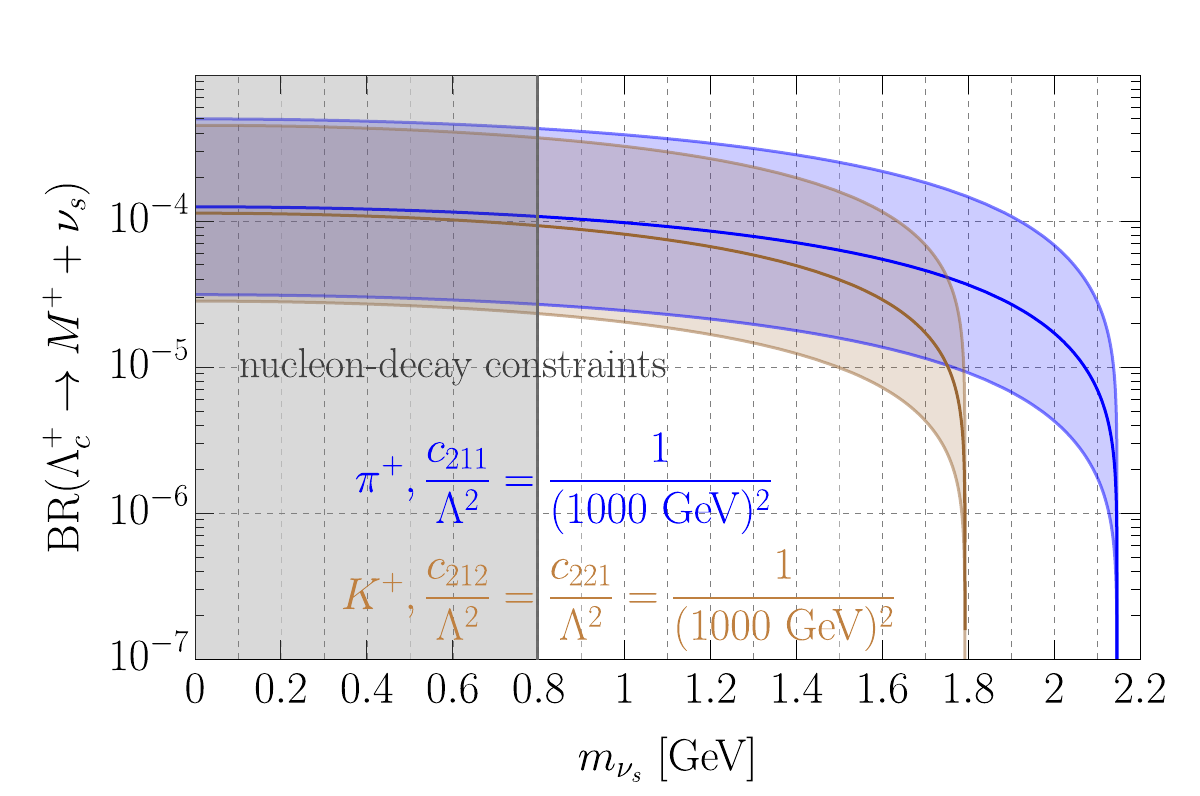}
      \caption{The decay branching ratios of $\Lambda_c^+\to M^+ + \nu_s$, where $M^+=\pi^+$ or $K^+$, as functions of $m_{\nu_s}$. Here, for the pion (kaon) case, $c_{211}/\Lambda^2$ ($c_{212}/\Lambda^2=c_{221}/\Lambda^2$) is fixed at $1/(1000\text{ GeV})^2$. The bands arise from varying the form factor $\beta_{\Lambda_c^+}$ by $2$ and $1/2$. The shaded region corresponds to $m_{\nu_s}<m_p-m_{\pi^+}$, where BNV nucleon-decay channels such as $p\to\pi^+ + \nu_s$ become kinematically allowed. Existing nucleon-decay searches provide extremely stringent limits and are therefore expected to impose much stronger constraints than the collider observables considered in this work.}
      \label{fig:lambdac_br}
\end{figure} 

In Fig.~\ref{fig:lambdac_br} we show the decay branching ratios of $\Lambda_c^+\to \pi^+/K^+ + \nu_s$ as functions of $m_{\nu_s}$, where we have fixed $c_{211}/\Lambda^2=1/(1000~\text{GeV})^2$ and $c_{212}/\Lambda^2=c_{221}/\Lambda^2=1/(1000~\text{GeV})^2$, for the pion and kaon cases, respectively.
The shaded region in Fig.~\ref{fig:lambdac_br} corresponds to $m_{\nu_s}<m_p-m_{\pi^+}$, where BNV nucleon-decay channels such as $p\to\pi^+ + \nu_s$ become kinematically allowed.
Existing nucleon-decay searches are expected to provide substantially stronger constraints in this mass range than the collider observables considered in this work.
Given the uncertainty associated with extrapolating the hadronic form factor $\beta_{\Lambda_c^+}$ beyond the domain of BChPT, we vary it by factors of $2$ and $1/2$, leading to the bands shown in this plot.
A more rigorous determination of these matrix elements would require dedicated non-perturbative calculations, such as lattice QCD or improved QCD sum-rule analyses for charmed baryons, which is beyond the scope of this work.

\subsection{The RPV-SUSY}\label{subsec:rpv}

We work with the low-scale RPV-SUSY model with the hypothesis of single-coupling dominance, where we assume that only the coupling $\lambda''_{212}$ among all is non-vanishing.
We first present the parton-level RPV-SUSY Lagrangian $\mathcal{L}_{\text{BNV-bino}}$ relevant to our study which focuses on a pure bino neutralino associated with the BNV interactions induced by the single coupling $\lambda''_{212}$:\footnote{We note that for the $\Lambda_c^+\to \pi^++\tilde{\chi}^0_1$ process the RPV coupling $\lambda''_{211}$ would be required, which is, however, vanishing in the RPV-SUSY. Therefore, for the case of the RPV-SUSY, we restrict ourselves to the $\Lambda_c^+\to K^+ + \tilde{\chi}^0_1$ decay which is mediated by the coupling $\lambda''_{212}$.}~\cite{Weinberg:1981wj,Hall:1983id,Faessler:2007br,Li:2007ih}
\begin{eqnarray}
\mathcal{L}_{\text{BNV-bino}}&=& \mathcal{L}_{\text{bino}}+ \mathcal{L}_{\text{RPV}}, \label{eq:Lagrangian-RPV-SUSY}\\
\mathcal{L}_{\text{bino}}&=& - \sum\limits_{q=d,s,c}  g^{\tilde q}_{1R} \left(\bar q_{R,a} P_L \tilde{\chi}^0_1\right) \tilde q_{R,a} + {\text{h.c.}} + \ldots, \label{eq:Lagrangian-bino} \\
\mathcal{L}_{\text{RPV}} &=&  \lambda''_{212}  \, \epsilon_{abc}   \left(          \tilde c^{*}_{Ra}\, \bar d_{Rb}\, s^{C}_{Rc}            + \tilde d^{*}_{Ra}\,  \bar c_{Rb}\,  s^{C}_{Rc}      + \tilde s^{*}_{Ra}\,  \bar c_{Rb}\,  d^{C}_{Rc}         \right) + {\text{h.c.}}, \label{eq:Lagrangian-RPV}
\end{eqnarray}
where $\mathcal{L}_{\text{bino}}$ contains the pure bino interaction terms with a quark and a squark (we only include the $d, s,$ and $c$ quark flavors since we restrict ourselves to the $\lambda''_{212}$ coupling).
$g^{\tilde q}_{1R}$ is given as follows,
\begin{eqnarray}
g^{\tilde q}_{1R} = - \sqrt{2} \, g_W\,  e_q  \tan\theta_W,\label{eqn:bino_coupling}
\end{eqnarray}
with $\tan{\theta_W}\simeq 0.54840$ for the weak mixing angle $\theta_W$, $g_W=e/\sin{}\theta_W\simeq 0.62977$ being the $SU(2)$ gauge coupling, and $e_q$ labeling the charge of the associated quark such that $e_c=2/3$ and $e_{d/s}=-1/3$.
$P_L= (1-\gamma^5)/2$ is the usual left-chiral projector.
Further, $a, b,$ and $c$ in Eq.~\eqref{eq:Lagrangian-RPV} are quark color indices, with $\epsilon_{abc}$ being the anti-symmetric Levi-Civita tensor.
The upper index $C$ labels charge conjugation.

In Eq.~\eqref{eq:Lagrangian-RPV}, an overall factor $1/2$ usually included in the literature is omitted, since the coupling $\lambda''_{221}=-\lambda''_{212}$ is absorbed here into the definition of $\lambda''_{212}$.
We note that in Eq.~\eqref{eq:Lagrangian-bino}, only terms with right-chiral quarks or squarks are given, since we assume absence of squark mixing and the BNV terms as shown in Eq.~\eqref{eq:Lagrangian-RPV} involve only such fields.
Besides assuming vanishing squark mixing, we confine ourselves to the case of degenerate squark masses $m_{\tilde{q}}$ for simplicity of discussion.
\begin{figure}[t]
      \centering
      \includegraphics[width=0.49\linewidth]{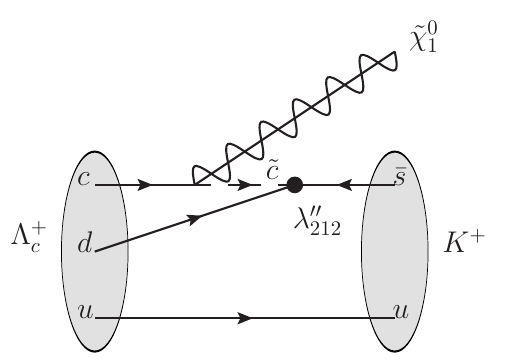}
      \includegraphics[width=0.49\linewidth]{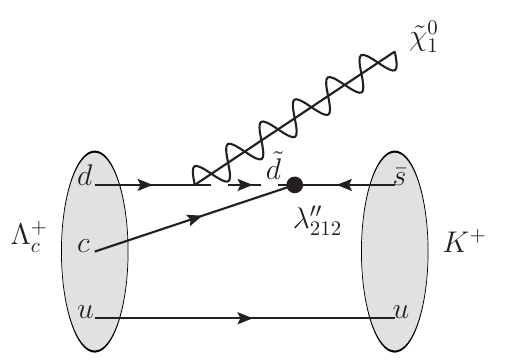}
      \includegraphics[width=0.49\linewidth]{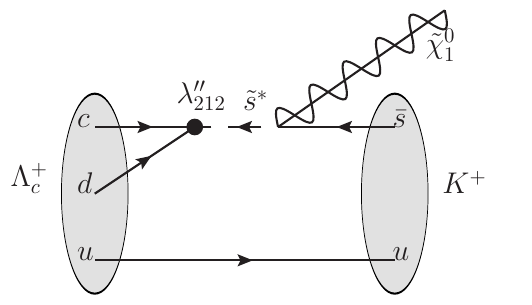}
      \caption{The parton-level Feynman diagrams for the $\Lambda_c^+\to K^++\tilde{\chi}^0_1$ decay process induced by the RPV coupling $\lambda''_{212}$.}
      \label{fig:feynman_diagrams_RPV}
\end{figure} 
The corresponding parton-level Feynman diagrams for the $\Lambda_c^+\to K^++\tilde{\chi}^0_1$ process are displayed in Fig.~\ref{fig:feynman_diagrams_RPV}.

Since the squark masses are expected to be much higher than the meson-mass scales, it is legitimate to integrate out the squarks in the theory, allowing us to write down the following dim-6 effective Lagrangian~\cite{Dib:2022ppx} for the BNV-process $\Lambda_c^+\to K^+ + \tilde{\chi}^0_1$:
\begin{eqnarray}
{\mathcal L}^{\text{BNV}} &=& {\mathcal L}^{scd\tilde{\chi}^0_1} + {\mathcal L}^{cds\tilde{\chi}^0_1}  +  {\mathcal L}^{dcs\tilde{\chi}^0_1} ,   \label{eq:Lagrangian_effetive_BNV}
\end{eqnarray}
where
\begin{eqnarray}
{\mathcal L}^{q_1q_2q_3\tilde{\chi}^0_1} &=& {\mathcal O}^{q_1q_2q_3} \,  \tilde{\chi}^0_1 + \text{h.c.}.
\end{eqnarray}
Here, ${\mathcal O}^{q_1q_2q_3}$ is defined as
\begin{eqnarray}
{\mathcal O}^{q_1q_2q_3} &=&  g^{\tilde q_1R}  {\mathcal O}_{q_1q_2q_3}^{LL},
\end{eqnarray}
with
\begin{eqnarray}
g^{\tilde qR} &=& \frac{g^{\tilde q}_{1R} \lambda''_{212}}{m_{\tilde q}^2},\\
{\mathcal O}_{q_1q_2q_3}^{LL} &=& \varepsilon_{a b c}   \left(\bar q_{3,c} P_L C \bar q_{2,b}^T\right) \bar q_{1,a} P_L ,
\end{eqnarray}
where $C=i \gamma^0 \gamma^2$ denotes the charge-conjugation matrix.

We thus arrive at the following effective Lagrangian,
\begin{eqnarray}
    \mathcal{L}^{\text{BNV}}=
    \frac{g^{\tilde{s}}_{1R}\lambda''_{212}}{m^2_{\tilde{s}}}(\overline{d_R}c^c_R)(\overline{s_R}\tilde{\chi}^0_1)
    +\frac{g^{\tilde{d}}_{1R}\lambda''_{212}}{m^2_{\tilde{d}}}(\overline{s_R}c^c_R)(\overline{d_R}\tilde{\chi}^0_1)
    +\frac{g^{\tilde{c}}_{1R}\lambda''_{212}}{m^2_{\tilde{c}}}(\overline{s_R}d^c_R)(\overline{c_R}\tilde{\chi}^0_1),
\end{eqnarray}
where the first two terms entail operators that match with the Hermitian-conjugated (h.c.) counterparts of the operators given in Eq.~\eqref{eqn:ef_op_cds} and Eq.~\eqref{eqn:ef_op_csd}, respectively.
The operator in the third term can be re-expressed via a Fierz transformation as
\begin{eqnarray}
    (\overline{s_R}d^c_R)(\overline{c_R}\tilde{\chi}^0_1)\overset{\text{Fierz}}{\approx}\frac{1}{2}(\overline{s_R}\tilde{\chi}^0_1)(\overline{c_R}d^c_R)=-\frac{1}{2}(\overline{d_R}c^c_R)(\overline{s_R}\tilde{\chi}^0_1),\label{eqn:operator_Fierz}
\end{eqnarray}
where an extra tensor term that originates from the Fierz transformation and is expected to give negligible contributions, has been ignored~\cite{Gunther:2023vmz}.
The last expression in Eq.~\eqref{eqn:operator_Fierz} matches the h.c.~counterpart of Eq.~\eqref{eqn:ef_op_cds}.
Therefore, we arrive at the following expression,
\begin{eqnarray}
    \mathcal{L}^{\text{BNV}}=
    2\frac{g^{\tilde{s}}_{1R}\lambda''_{212}}{m^2_{\tilde{q}}}(\overline{d_R}c^c_R)(\overline{s_R}\tilde{\chi}^0_1)
    +\frac{g^{\tilde{d}}_{1R}\lambda''_{212}}{m^2_{\tilde{q}}}(\overline{s_R}c^c_R)(\overline{d_R}\tilde{\chi}^0_1),
\end{eqnarray}
where we have assumed squark-mass degeneracy and exploited the relation $g_{1R}^{\tilde{c}}=-2 g_{1R}^{\tilde{s}}$ derived from Eq.~\eqref{eqn:bino_coupling}.

Matching the UV Lagrangian to the effective Lagrangian~\eqref{eqn:cds_eft_lag}, we obtain the following relations,
\begin{eqnarray}
    \frac{c_{212}}{\Lambda^2}=\frac{2\lambda''_{212}g^{\tilde{s}}_{1R}}{m_{\tilde{q}}^2},
    \quad
    \frac{c_{221}}{\Lambda^2}=\frac{\lambda''_{212}g^{\tilde{d}}_{1R}}{m^2_{\tilde{q}}},
\end{eqnarray}
where $c_{212}/\Lambda^2$ and $c_{221}/\Lambda^2$ are, respectively, the Wilson coefficients of the $\mathcal{O}^{S,RR}_{cds}$ and $\mathcal{O}^{S,RR}_{csd}$ EFT operators, rescaled by the NP scale $\Lambda$ squared.

\subsection{Decays of the sterile neutrino and the light bino neutralino}\label{subsec:decay_llp}

Before closing the section, we briefly comment on the decay length of $\nu_s$ and $\tilde{\chi}^0_1$.
For the considered mass range of interest, their decays are highly suppressed.
For the sterile neutrino, the decay amplitudes are proportional to the small values of the coefficients $c/\Lambda^2$ and are further reduced by off-shell propagators (a $W$-boson and a down-type quark).
Similarly, the light bino neutralino has its decays suppressed by three off-shell propagators (a squark, a $W$-boson, and a down-type quark), CKM matrix elements, tiny RPV couplings, as well as the absence of squark mixing.
Therefore, we conclude that both the sterile neutrino and light bino neutralino in the mass range of our interest have so long lifetimes that they appear as missing energy in the STCF main detector.
See also the numerical analyses in Ref.~\cite{Hiller:2026osz}.

\section{STCF and search analyses}\label{sec:exp}

We consider the energy-symmetric $e^+ e^-$ collider, STCF, operated at the COM energy $\sqrt{s}=4.682$~GeV, leading to large samples of $\Lambda_c^+ \overline{\Lambda}_c^-$ pair production events: $N_{\Lambda_c^+\overline{\Lambda}_c^-}=1.881\times 10^8$ per year with an integrated luminosity of 1~ab$^{-1}$~\cite{BESIII:2023rwv,Achasov:2023gey}.
With the MC simulation tool \textsc{OSCAR} developed for STCF, we generate signal events, perform tagging and event-reconstruction, and impose selection cuts.
\textsc{OSCAR} is built on the \textsc{SNiPER} architecture, utilizing \textsc{DD4hep} for detailed detector description and \textsc{PODIO} for data modeling, thus providing a modular and efficient platform for large-scale MC sample production~\cite{Huang:2022bkz,Xiang:2023mkc,Zeng:2023wqw}. 

In the signal processes, we start with simulating $e^+ e^- \to \Lambda_c^+ \overline{\Lambda}_c^-$.
The $\overline{\Lambda}_c^-$ baryon should be tagged, based on its different decay modes.
Normally 12 tag channels are selected for analysis~\cite{BESIII:2026qbp},
whereas we focus on the dominant tag channel,  $\overline{\Lambda}_c^- \to \overline{p} K^+ \pi^-$ of which the branching ratio is $6.35\%$~\cite{ParticleDataGroup:2024cfk}, for estimating the reconstruction efficiencies of the signal events.
Including into the analysis the other tag channels would enhance the signal-event rates by a factor of $\mathcal{O}(1)$, since the used tag channel $\overline{\Lambda}_c^- \to \overline{p} K^+ \pi^-$ constitutes roughly $40\%$ of all tag channels' BR sum.
The $\Lambda_c^+$ baryon should undergo the signal decays into a charged pion or kaon, plus a light NP particle being either $\nu_s$ or $\tilde{\chi}^0_1$.

In \textsc{OSCAR}, physics processes such as initial state radiation and beam energy spread are simulated using the \textsc{KKMC} generator~\cite{Jadach:2000ir,Jadach:1999vf}.
Decay modeling is handled by \textsc{EvtGen}~\cite{Lange:2001uf,Ping:2008zz} and final state radiation is implemented using \textsc{PHOTOS}~\cite{Barberio:1990ms}.
We use the \textsc{PHOTOS} \textsc{PHSP} module to handle the $\Lambda_c^+$ signal decay's kinematics including the angular distributions of the final-state particles.\footnote{Strictly speaking, the decay-product distribution of $\Lambda_c^+$ should have a minor dependence on the polar angle. Here, for simplicity, we adopt a pure phase-space model for the $\Lambda_c^+$ decays. The associated modeling uncertainty is expected to be subdominant compared to the uncertainties arising from the heavy-flavor-baryon extrapolation discussed above.}
The tool \textsc{OSCAR} takes into account the efficiencies of reconstructing both the tagged $\overline{\Lambda}_c^-$ baryon and the signal meson, which depend on the mass of the light NP particle.

For signal events, we first reconstruct charged-particle tracks.
We start with extracting the likelihood value of each particle for particle identification: if the kaon likelihood value is both positive and larger than those of the pion and the proton, the particle is identified as a kaon. The same criterion applies to the identification of the proton and pion.

After particle identification, we extract the distances of closest approach of each track to the interaction point, defined as $|V_z|$ along the beam ($z$) direction and $|V_{xy}|$ in the plane transverse to the beam. 
We require $|V_z| < 10.0$~cm and $|V_{xy}| < 1.0$~cm.
The polar angle $\theta$ of each track with respect to the beam axis is required to satisfy $|\cos\theta| < 0.93$.
These requirements are applied equally to the selection of protons, kaons, and pions.
The tracks satisfying all the criteria given above are considered as qualified.

Before proceeding to decay reconstruction, we apply a selection based on the number of qualified particles in the event: if the signal side selects a $K^+$ ($\pi^+$), it is required that $N(K)\ge 2,N(p)\ge1,N(\pi)\ge1$ ($N(K)\ge 1,N(p)\ge1,N(\pi)\ge2$).

When selecting qualified $\Lambda_c^+$ and $\overline{\Lambda}_c^-$ candidates, we use the double-tag method.
On the tag side, we first apply a charge condition: the proton and pion on the tag side are required to have the same charge, while the kaon and proton are required to have opposite charges.
After performing a kinematic fit on the candidate $\overline{\Lambda}_c^-$ on the tag side, the fitting $\chi^2$ is required to be positive.

The beam-constrained mass $m_{\rm BC}\equiv\sqrt{E_{\rm beam}^2-|\mathbf{p}_{\overline{\Lambda}_c^-}|^2}$, where $E_{\rm beam}$ and $\mathbf{p}_{\overline{\Lambda}_c^-}$ are respectively the beam energy and momentum of the selected $\overline{\Lambda}_c^-$ candidate in the COM system, is calculated for each candidate. 
The candidates lying within the range of $2.25$~GeV to $2.4$~GeV on the $m_{\rm BC}$ distribution are kept for further analysis.
Also, the energy difference $\Delta E \equiv E_{\bar\Lambda_c^-} - E_{\text{beam}}$ is calculated, where $E_{\overline{\Lambda}_c^-}$ is the total energy of the $\overline{\Lambda}_c^-$ candidate in the COM system.
$\Delta E$ is restricted to the range between $-29$~MeV and $26$~MeV.
In addition, the candidate with the smallest $\Delta E$ is selected as the tag-side candidate.

On the signal side with a $K^+$ ($\pi^+$), it is required that the charged kaon (pion) selected has the same charge as the kaon on the tag side.
Finally, the missing mass squared $m_{\text{missing}}^2$ is required to be positive.

\begin{figure}[t]
      \centering
    \includegraphics[width=0.495\textwidth, height=7cm]{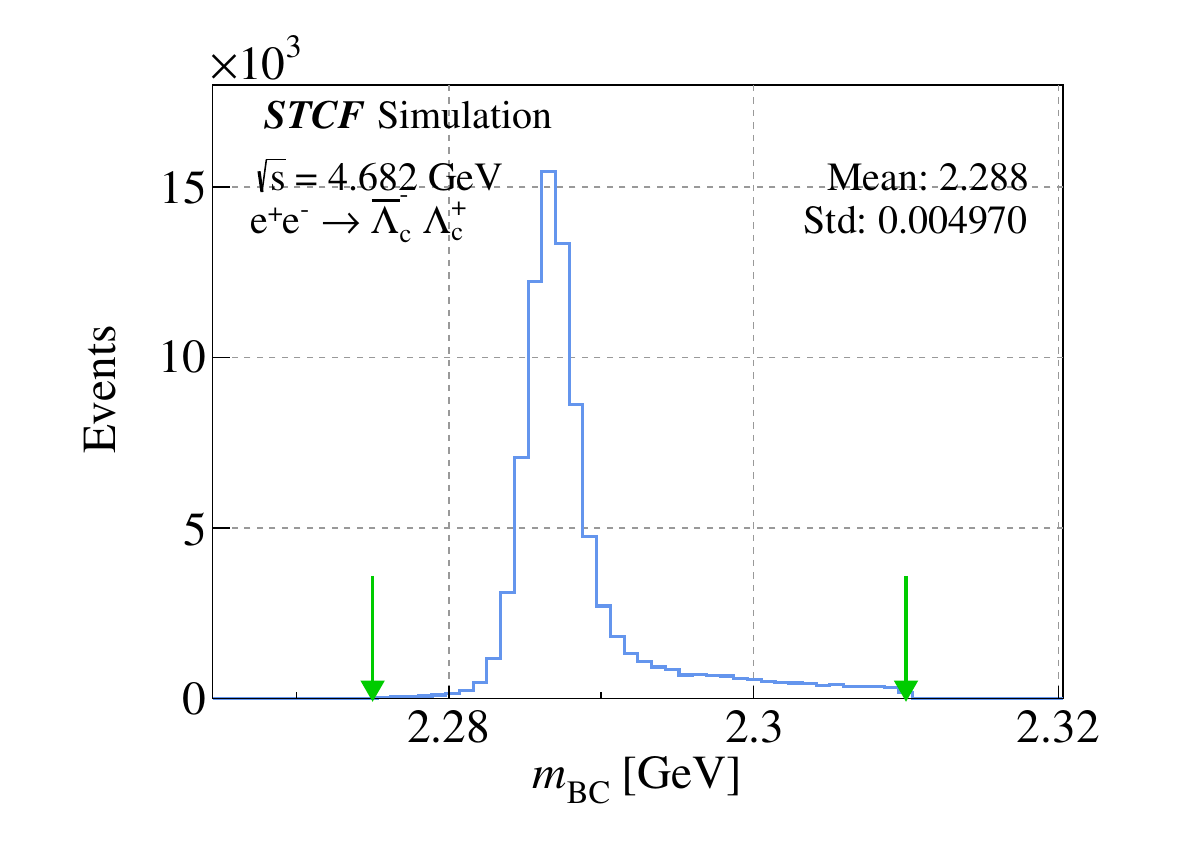}
    \includegraphics[width=0.495\textwidth, height=7cm]{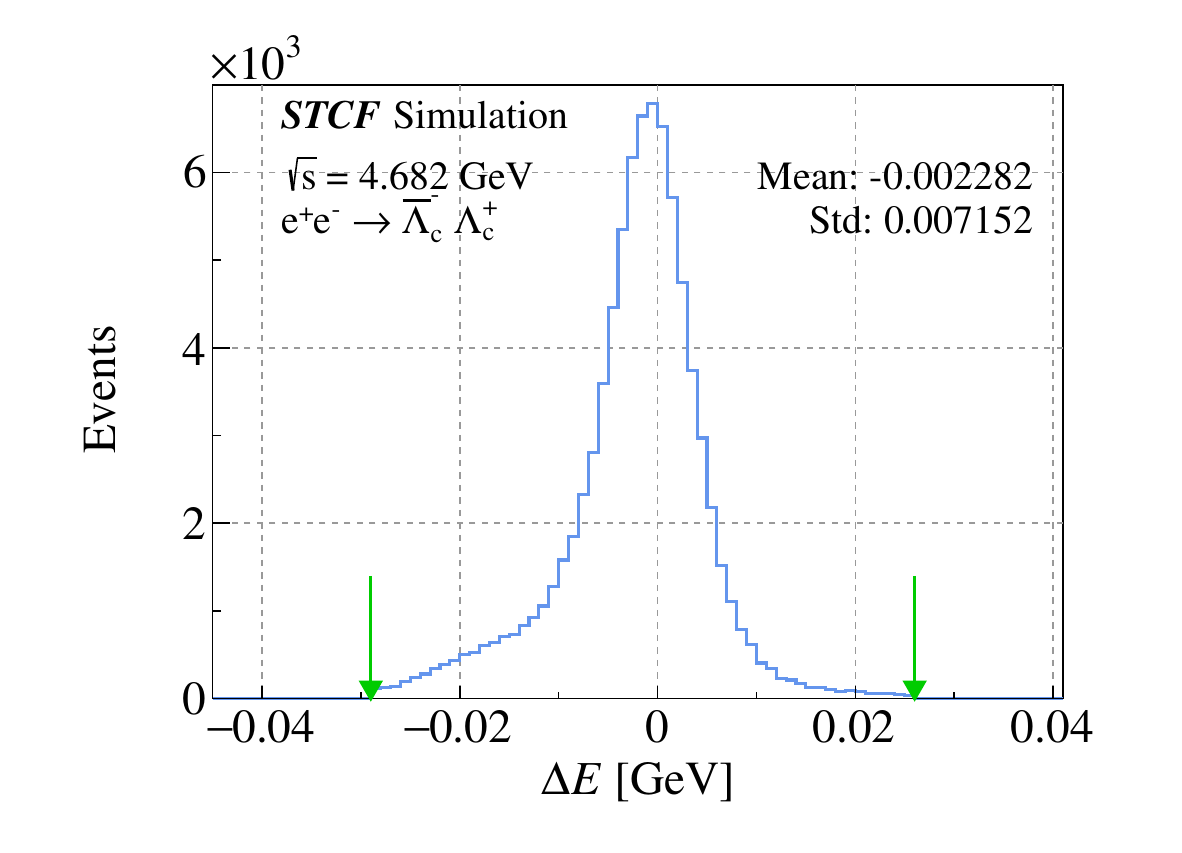}
    \caption{The distributions of $m_{\text{BC}}$ (left) and $\Delta E$ (right) for $m_{\text{missing}}=1.110$~GeV, at the COM energy $\sqrt{s}=4.682$~GeV. The downward arrows correspond to the event-selection windows we adopt. A total of twenty thousand events are simulated. On both plots, the mean values and the standard deviations are given. On the signal side, the $\Lambda^+_c$ baryon is assumed to decay to $K^++$~missing; however, in the pion case the distribution shape would be almost identical.}
    \label{fig:DeltaE_mBC}
\end{figure}

We show in Fig.~\ref{fig:DeltaE_mBC} distributions of $m_{\text{BC}}$ and $\Delta E$ for the $\overline{\Lambda}^-_c$ candidates, for a missing mass of 1.110~GeV, where the $\Lambda^+_c$ baryon is assumed to undergo a signal decay into $K^++$~missing.
The downward pointing arrows correspond to the selection windows on $\Delta E$ and $m_{\text{BC}}$.
In total, twenty thousand events are simulated.
We note that both distributions have a negligible dependence on the missing mass, unless the missing mass is close to the kinematic thresholds.

These numerical simulations provide the reconstruction efficiencies of the signal events.
We then compute the signal-event numbers with
\begin{eqnarray}
    N_S=2\cdot N_{\Lambda_c^+\overline{\Lambda}_c^-}\cdot \text{BR}(\Lambda_c^+\to M^+ + \nu_s/\tilde{\chi}^0_1)\cdot \text{BR}(\overline{\Lambda}^-_c\to \overline{p} K^+ \pi^-)\cdot  \epsilon,  \label{eqn:NS}
\end{eqnarray}
where $M^+=(\pi^+, K^+)$, the factor $2$ accounts for the fact that there are two $\Lambda_c$ baryons in each signal event, and $\epsilon$ is the mass-dependent reconstruction efficiency obtained with \textsc{OSCAR} simulations.

Under the assumption of negligible background, $N_S=3$ corresponds to the exclusion limit at $95\%$ confidence level (C.L.).

A brief discussion of potential background sources is in order.
The dominant backgrounds are expected to arise from neutron-induced activities and continuum $e^+e^-\to$ light-quark processes, which can deposit energy in the electromagnetic calorimeter and mimic missing-energy signatures.
In a realistic analysis, such backgrounds can be substantially suppressed using optimized event selections, potentially assisted by advanced machine-learning techniques, while maintaining high signal efficiency~\cite{BESIII:2024cbr,BESIII:2024mgg}.
A detailed evaluation of these effects is left for future experimental studies.

\section{Numerical results}\label{sec:results}

\begin{figure}[t]
      \centering
      \includegraphics[width=0.495\linewidth]{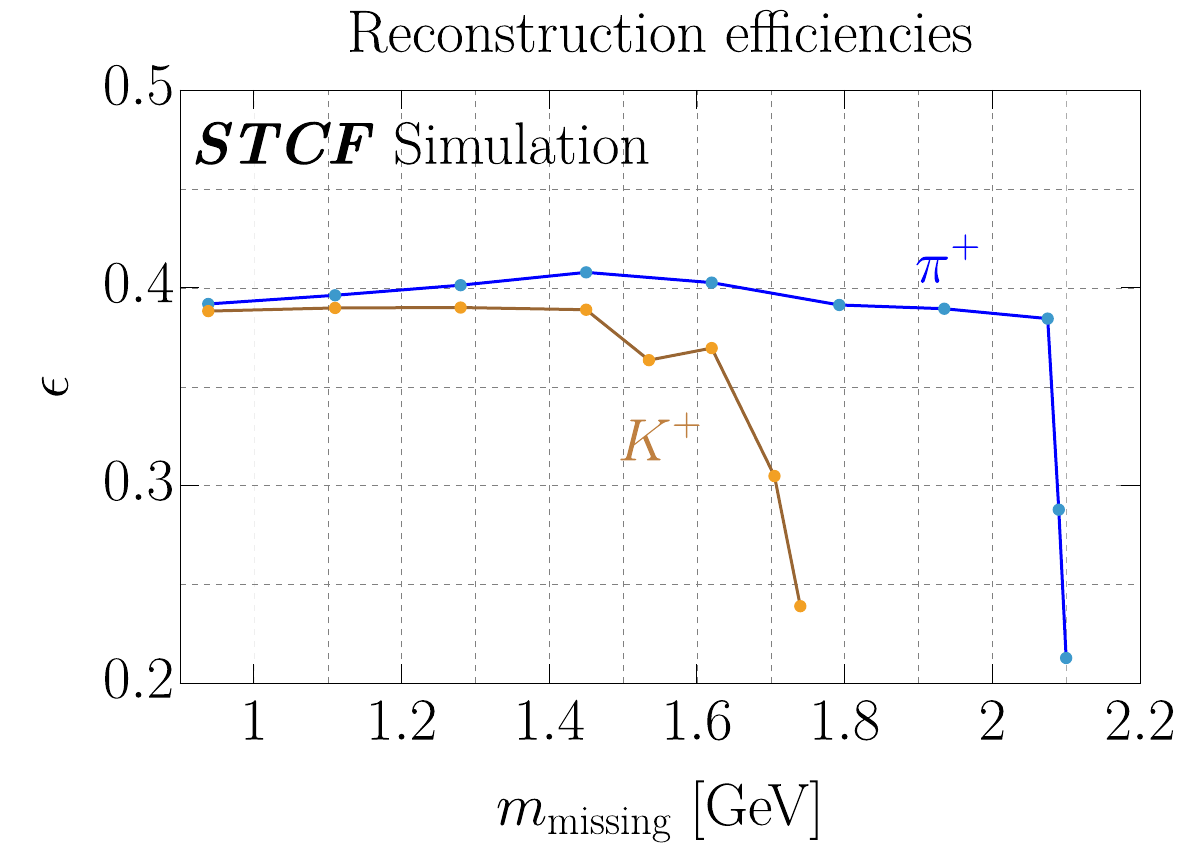}
      \includegraphics[width=0.495\linewidth]{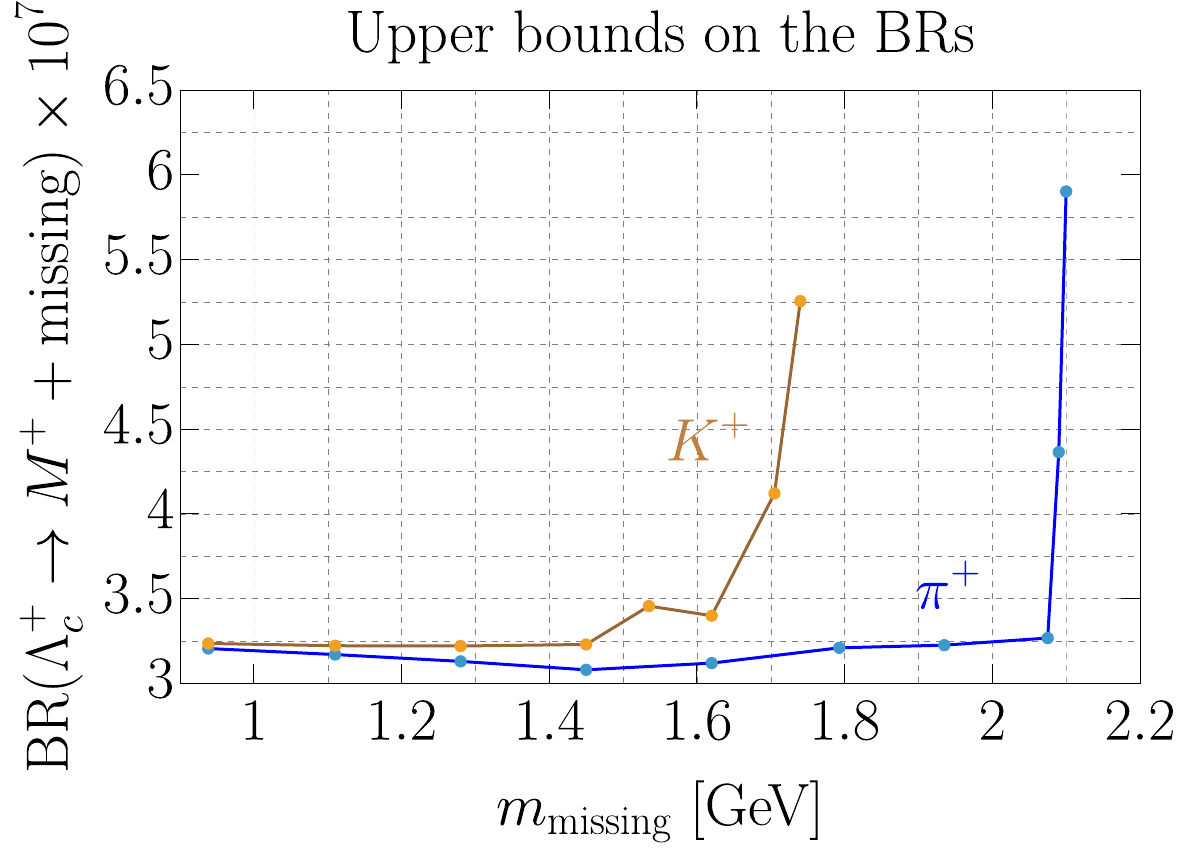}
      \caption{Left panel: the reconstruction efficiencies $\epsilon$ vs.~$m_{\text{missing}}$ for the pion (blue) and kaon (brown) channels.
      Right panel: model-independent upper bounds on BR$(\Lambda_c^+ \to \pi^+/K^++\text{missing})$ as functions of the missing mass $m_{\text{missing}}$, for vanishing background levels and an integrated luminosity of 1~ab$^{-1}$.}
      \label{fig:sensitivity_model_indep}
\end{figure}

In the left plot of Fig.~\ref{fig:sensitivity_model_indep} we present the reconstruction efficiencies of the signal events as functions of the missing-energy invariant mass, for the pion (blue) and kaon (brown) channels.
We observe that the efficiencies are all close to 40\% for small values of $m_{\text{missing}}$, and drop quickly once the missing-energy mass approaches the kinematic thresholds.
In the right panel, we display the corresponding model-independent sensitivity results at $95\%$ C.L.~in the $\text{BR}(\Lambda_c^+\to M^++\text{missing})$~vs.~$m_\text{missing}$ plane, derived from Eq.~\eqref{eqn:NS} by requiring $N_S=3$.
We observe that STCF can probe BR$(\Lambda_c^+\to M^++\text{missing})$ as low as $\mathcal{O}(10^{-7})$.

\begin{figure}[t]
      \centering
      \includegraphics[width=0.49\linewidth]{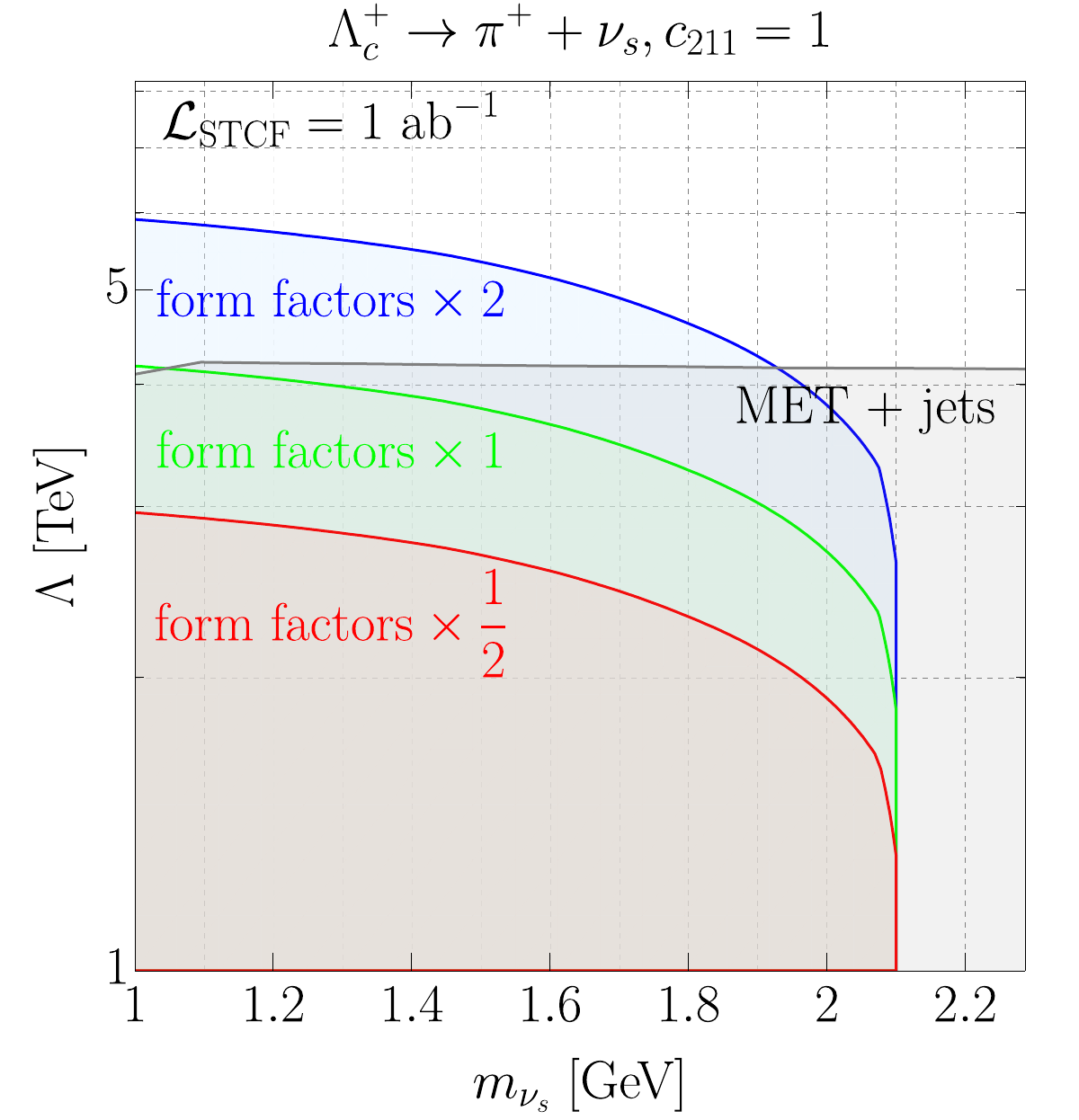}
      \includegraphics[width=0.49\linewidth]{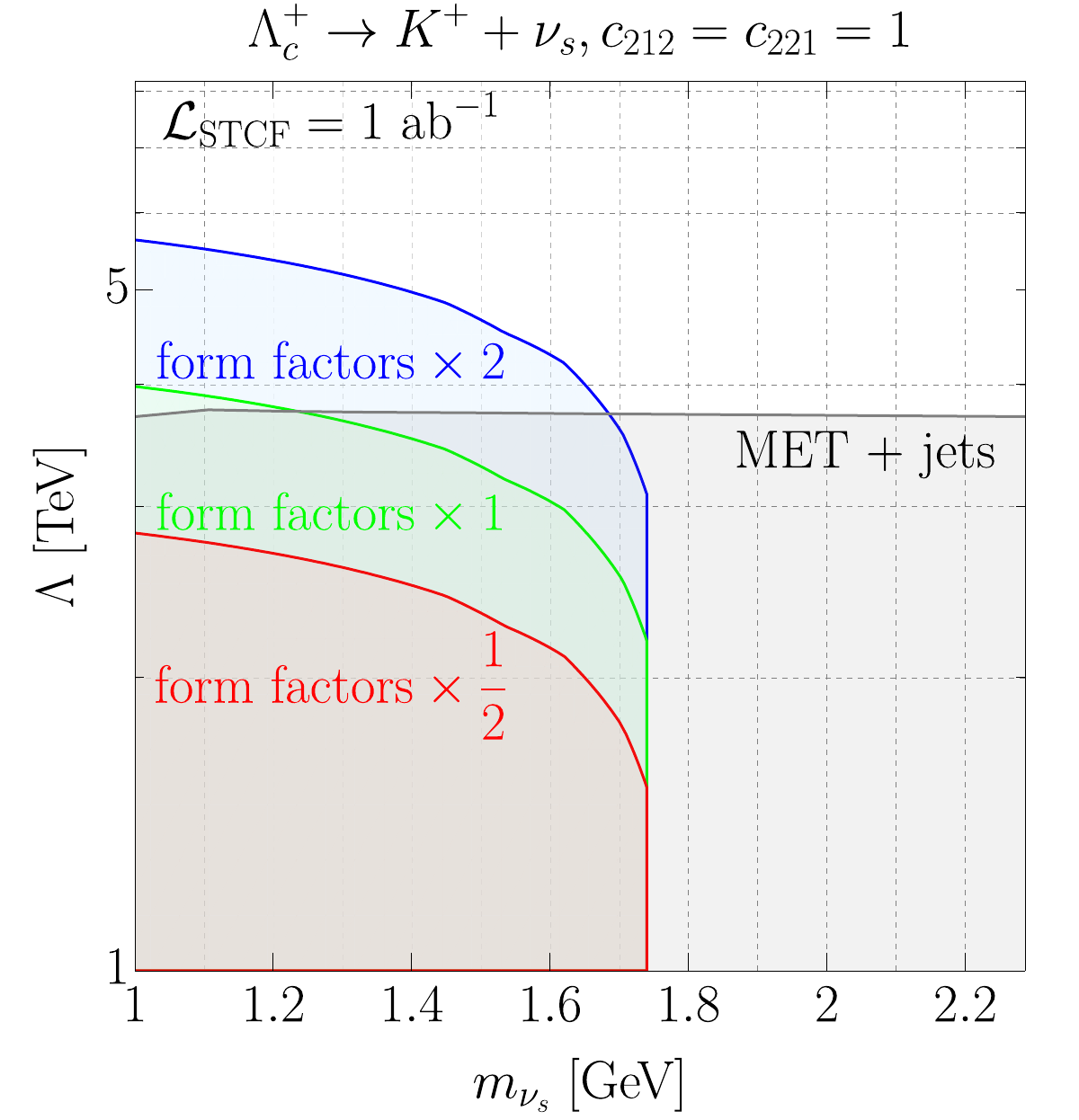}
      \includegraphics[width=0.49\linewidth]{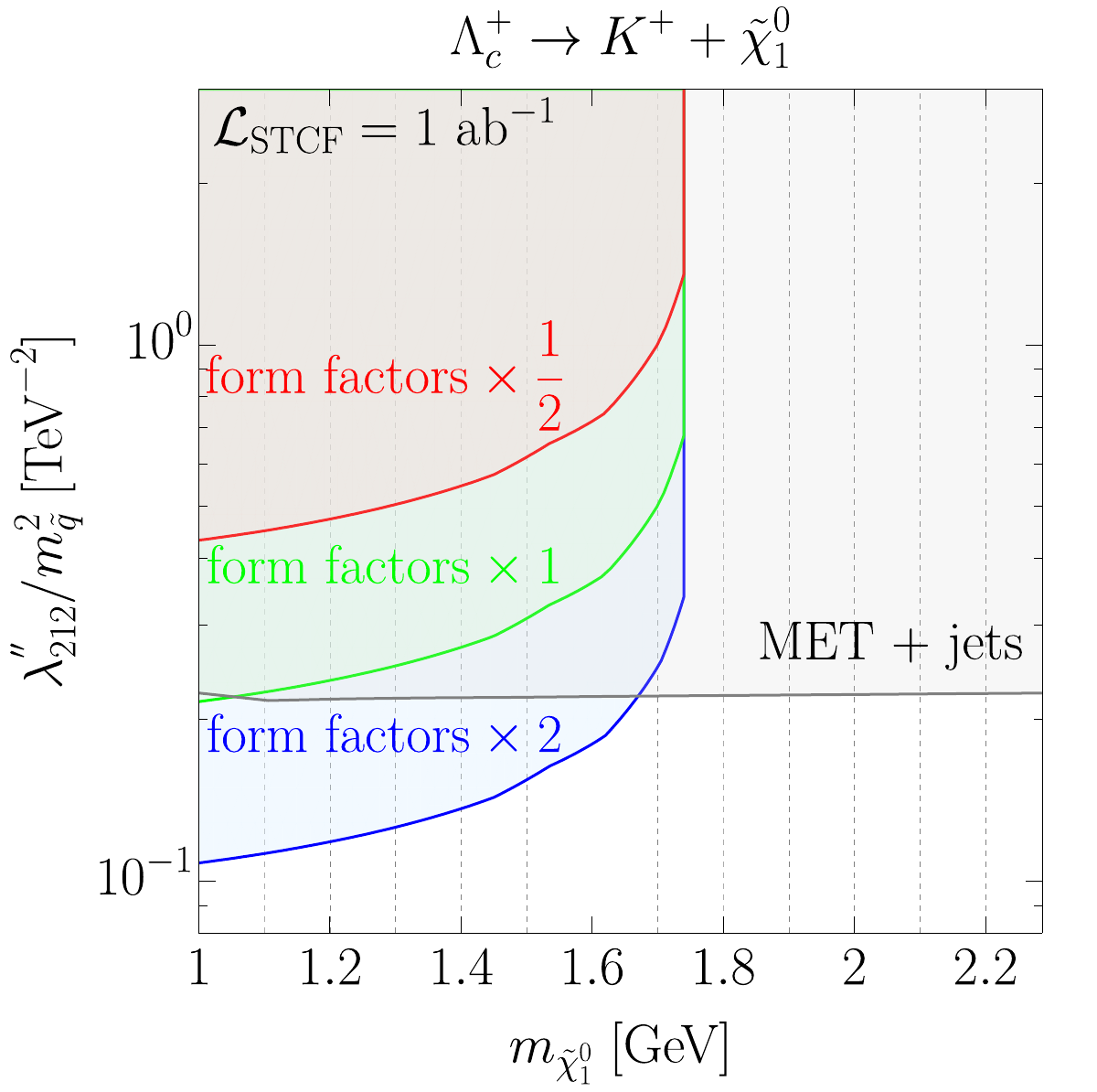}
      \caption{Model-dependent sensitivity limits at 95\%~C.L., for vanishing backgrounds. The leading LHC bounds on $\Lambda$ from the MET+jets search~\cite{ATLAS:2024vqf} are extracted from Ref.~\cite{Hiller:2026osz} and  we reinterpreted them in terms of the RPV parameters.}
      \label{fig:sensitivity_model_dep}
\end{figure}

We show in Fig.~\ref{fig:sensitivity_model_dep} model-dependent sensitivity results both within the $\nu$LEFT framework (the upper panels) and the RPV-SUSY model (the lower panel).
The upper plots correspond to $\Lambda_c^+ \to \pi^+ + \nu_s$ and $\Lambda_c^+ \to K^+ + \nu_s$, respectively.
Varying the hadronic form factor $\beta_{\Lambda_c^+}$ by factors of $2$ and $1/2$ to estimate the associated theoretical uncertainty from the extrapolation, we find that STCF is expected to probe NP scales $\Lambda$ up to roughly between $3$~TeV and $6$~TeV, assuming benchmark values $c_{211}=1$ and $c_{212}=c_{221}=1$, where $\Lambda$ should be interpreted as the effective scale suppressing the dimension-6 operators.
For other choices of the Wilson coefficients, the corresponding reach can be trivially rescaled as $\Lambda \propto \sqrt{c}$.
For instance, taking $c \sim 4\pi$, as motivated by strongly coupled scenarios, would correspond to an increase of the inferred scale $\Lambda$ by a factor of $\sim \sqrt{4\pi} \simeq 3.5$.

The existing leading bounds stem primarily from LHC searches for the signatures of transverse missing energy (MET)+jets~\cite{ATLAS:2024vqf} and MET+top~\cite{ATLAS:2020yzc}.
Recently, the authors of Ref.~\cite{Hiller:2026osz} have recast these searches and reinterpreted the constraints in terms of the operators we study here.
For the mass range of our interest, the search for MET+jets with an integrated luminosity of 140~fb$^{-1}$~\cite{ATLAS:2024vqf} turns out to always dominate, for which the constraints on $\Lambda$ roughly range between 3.5~TeV and 4~TeV.
We note that in the kaon scenario, both $c_{212}$ and $c_{221}$ are assumed to be non-zero and equal in this work; Ref.~\cite{Hiller:2026osz} derives stronger bounds on $c_{212}$ than $c_{221}$ and therefore in the upper right panel we display the MET+jets bounds on $\Lambda$ for $c_{212}=1$.\footnote{When both $c_{212}$ and $c_{221}$ are present, interference effects can arise. However, as shown in Ref.~\cite{Hiller:2026osz}, two-dimensional fits in the $(c_{212}/\Lambda^2, c_{221}/\Lambda^2)$ plane yield bounds comparable to the single-coupling case (dominated by MET+jets searches), and are therefore not included here.}

Following Sec.~\ref{subsec:rpv} we proceed to reinterpret these exclusion bounds on $\Lambda$ in terms of the RPV-SUSY parameters, and display the projected sensitivity reach in the $(m_{\tilde{\chi}^0_1}, \lambda''_{212}/m_{\tilde{q}}^2)$ plane.
Our numerical findings indicate that for $m_{\tilde{\chi}^0_1}$ between 1~GeV and $m_{\Lambda_c^+}-m_{K^+}$, STCF is projected to be sensitive to the RPV model parameter $\lambda''_{212}/m^2_{\tilde{q}}$ down to $\mathcal{O}(0.1)~\mathrm{TeV}^{-2}$.

Finally, we note that for the theoretical scenarios studied above, besides the existing bounds from the proton lifetime and the LHC searches for MET+jets and MET+top, we may take into account constraints from dinucleon decays~\cite{Dib:2022ppx}; the corresponding bounds are, however, found to be inferior compared to the LHC bounds.
Therefore, we choose not to overlap the sensitivity plots with these bounds.

\section{Conclusions}\label{sec:conclusions}

In this work we propose to search for $\Lambda_c^+\to \pi^+/K^+ +\text{missing}$ with apparent BNV, which is predicted in a wide range of BSM scenarios, at the proposed STCF in China.
We focus on the operation mode of a COM energy $\sqrt{s}=4.682$~GeV leading to large rates of $\Lambda_c^+ \overline{\Lambda}_c^-$-pair near-threshold-production events.
We perform MC simulations with the STCF-dedicated tool \textsc{OSCAR} to determine the reconstruction efficiencies of the signal events.
Here, we have identified twelve tag channels, and confined ourselves to the dominant one, $\overline{p}K^+\pi^-$, for numerical analysis, considering its relatively large decay BR and excellent reconstruction efficiencies.

Our sensitivity estimates are obtained under the assumption of negligible background. Given the clean environment of near $\Lambda_c^+\overline{\Lambda}_c^-$ threshold production, residual background contributions are expected to be controllable and not to qualitatively affect our conclusions, although a dedicated experimental study is required for a quantitative assessment.

Given the signal-event reconstruction efficiencies with the $\overline{p}K^+\pi^-$ tag channel obtained from simulation, we have derived the projected upper limits on BR$(\Lambda_c^+\to \pi^+/K^++\text{missing})$ as functions of $m_{\text{missing}}$, for an integrated luminosity of 1~ab$^{-1}$.
These results indicate that our proposed searches can test BR$(\Lambda_c^+\to \pi^+/K^++\text{missing})$ down to $\mathcal{O}(10^{-7})$.
We have further converted these observable-level bounds to lower bounds on the NP scale $\Lambda$ as functions of $m_{\nu_s}$ and upper bounds on the RPV parameter combination $\lambda''_{212}/m^2_{\tilde{q}}$ as functions of $m_{\tilde{\chi}^0_1}$.
For this computation, we have promoted the hadronic form factors from the nucleon sector to the $\Lambda_c$ baryon, which lies outside the formal domain of applicability of BChPT, and treat the form factor $\beta_{\Lambda_c}$ as an effective parameter with an associated order-unity uncertainty.
The numerical results show that our proposed analyses can probe NP scales up to about $3$--$6$~TeV for $c=1$ and small values of $m_{\nu_s}$, depending mainly on theoretical uncertainties.
In terms of the model parameters in the RPV-SUSY with a light bino neutralino, we find that STCF can probe $\lambda''_{212}/m^2_{\tilde{q}}$ down to $\sim 10^{-1}$ TeV$^{-2}$.

In general, our results show that the proposed search channels at STCF with an integrated luminosity of $1$~ab$^{-1}$ can probe modest regions of the parameter space beyond the current bounds.
In addition, if further tag channels can be included into the analysis, we expect a reach to the NP scales up to about 8~TeV can be achieved.

Finally, we emphasize that BESIII and the proposed STCF are currently the only facilities worldwide capable of producing heavy baryons, such as $\Lambda_c^+\overline{\Lambda}^-_c$, directly at or near the threshold.
This production mode provides a relatively clean environment in which the baryons are created nearly at rest and in correlated pairs, enabling high-efficiency tagging and precise kinematic closure, including for final states with missing energy.
These features make threshold charm-baryon factories uniquely suited for probing apparently baryon-number-violating, semi-invisible decay modes.
Although other $e^+e^-$ colliders such as the FCC-ee~\cite{FCC:2025lpp} and Belle~II~\cite{Belle-II:2010dht,Belle-II:2018jsg} can in principle also test the scenarios studied here (see Ref.~\cite{Hiller:2026osz}), it would be difficult for them to match this combination of cleanliness, kinematic control, and sensitivity, underscoring the distinctive discovery potential of BESIII and STCF for apparent BNV phenomena associated with long-lived particles.

\textbf{Note added:} During the final preparation of this manuscript, Ref.~\cite{Hiller:2026osz} appeared on the arXiv. The authors of Ref.~\cite{Hiller:2026osz} have derived the current constraints on both $udd \nu_s$- and $qqd \nu_s$-type operators from the LHC, while focusing on the $qqd\nu_s$ operators for phenomenological analyses at electron-positron colliders without detailed detector simulations. In the present work, we instead focus on the complementary $udd\nu_s$-type operators and perform dedicated detector-level simulations for the semi-invisible decay channels $\Lambda_c^+\to M^+ + \nu_s$ at STCF using OSCAR. The two classes of operators probe complementary realizations of BNV involving sterile neutrinos.

\section*{Acknowledgments}

We would like to thank (in alphabetical order) Tong Li, Valery Lyubovitskij, Nicol\'as Neill, Jin Sun, Arsenii Titov, Kechen Wang, Fanrong Xu, Wenbiao Yan, Chang-Yuan Yao, and Xingbo Yuan, for useful discussions.
This work was supported by the National Natural Science Foundation of China under grant Nos.~12475106 and 12505120 and National Key R\&D Program of China under Contract No.\ 2023YFA1606004, and the Fundamental Research Funds for the Central Universities under Grant No.~JZ2025HGTG0252.
X.Z.~acknowledges support from the CAS Youth Team Program under Contract No.~YSBR-101.

\vspace{1cm}

\bibliography{refs}

\end{document}